\documentclass[12pt]{article}

\begin{document}

\begin{center}

{\Large {Schwarzschild deformed supergravity background: possible geometry origin of fermion generations and mass hierarchy}}

\vspace{1,5cm}

{Boris L. Altshuler}\footnote{E-mail addresses: baltshuler@yandex.ru \,\,\,  \&  \,\,\,  altshul@lpi.ru}

\vspace{1cm}

{\it {Theoretical Physics Department, Lebedev Physical
Institute of the Russian Academy of Sciences, \\  53 Leninsky Prospect, Moscow, 119991, Russia}}

\vspace{1,5cm}

\end{center}

{\bf Abstract:} 
The problem of  fermion masses hierarchy in the Standard Model is considered on a toy model of a 10-dimensional space-time with a IIA supergravity type background. Dirac equation on this background, after compactification of extra 4- and 1-dimensional subspaces, gives the spectrum of Fermi fields which profiles in 5 dimensions and corresponding Higgs generated masses in 4 dimensions  depend on the eigenvalues of Dirac operator on the named compact subspaces. Schwarzschild Euclidean deformation of the supergravity throat with the "apple-shaped" conical singularity permits to leave only three non-divergent angular modes interpreted as three generations of the down-type quarks. Calculated ratio $m_{d} / m_{s} = e^{-3}$ exactly coincides with its experimentally observed value for integer values of two free parameters of the 10-dimensional background. Equations for non-chiral modes coincide with the non-relativistic Schrödinger equation for an electron moving in a Coulomb field; the corresponding small fermion masses generated by the twisted boundary conditions are expressed through the degenerate hypergeometric functions.

\vspace{0,5cm}


\vspace{0,5cm}


\newpage

\section{Introduction}

\quad Models in warped extra dimensions are among the popular trends aimed at resolving the Standar Model (SM) fermions mass hierarchy enigma. In the works of this approach, based upon $5D$ Randall-Sundrum (RS) model \cite{RS}, chiral SM fermions "live" in the bulk and their small masses $m_{f}$ in 4 dimensions result from the overlaps of the fermionic bulk wave functions and the Higgs field profile confined to the IR end of the slice of the RS $AdS_{5}$ background, see reviews \cite {5D1} - \cite{5D3} and references therein. Then the calculated values of masses of quarks or leptons strongly depend on the ratios $c_{f} = M_{f} / k$ of bulk masses $M_{f}$ of corresponding Fermi fields to the curvature $k$ of $AdS_{5}$: $m_{f} \sim \epsilon^{c_{f} - 1/2}$, where $\epsilon = 10^{-16}$ is the Planck-TeV hierarchy parameter. Thus to get the observed masses $m_{f}$ the special choice of values of parameters $c_{f}$ in vicinity of 1/2 for every SM fermion is demanded. This fine-tuning of fermion bulk masses is an essential drawback of models \cite{5D1} - \cite{5D3}. We also pay attention to papers \cite{Fujimoto}, \cite{Isidori1} where each Standard Model family "lives" in its own slice of the $5D$ bulk separated by additional 3-branes and fermion masses in 4 dimensions essentially depend on the choice of parameters of the model, and to paper \cite{Neubert}, where fermions bulk masses are dynamically generated through fermions interaction with the bulk scalar field, but again bulk Yukawa coupling constants of this interaction are fine-tuned for every SM fermion.

In the present paper the Dirac equation with zero bulk mass on a ten-dimensional background is considered where profiles of Fermi fields in the $5 D$ space-time remaining after compactification of (4 + 1) extra dimensions are determined by eigenvalues $K^{(4)}_{l}$ of Dirac operator on compact subspace $K^{(4)}$ and by the angular momentum $q$ with respect to the warped $S^{1}$ dimension. In this way the present paper generalizes the approach of \cite{Neronov} - \cite{Libanov} where the named angular momentum numbers the fermion generations and three fermion families arise from a single Fermi field in six dimensions. In the $6D$ model of \cite{Apple} three fermion generations originate thanks to the introduced codimension two brane and corresponding "apple-shaped" conical singularity of the extra two-dimensional sphere. We use the same trick at the point of the Schwarzschild Euclidean "horizon" of the supergravity throat.

Beginning from Witten's seminal work \cite{Witten} it is known that the chiral fermions of SM cannot be obtained from Dirac equation on $D$-dimensional space $M_{4} \times \Sigma_{D-4}$ purely from the metric and spin connection on compact extra space $\Sigma_{D-4}$. In order for the Dirac operator on extra space to have physically necessary zero modes the topologically non-trivial background gauge fields must be introduced, as it is done, for example, in \cite{Dolan1} - \cite{Dvali}. However, as it is shown below, Witten’s no-go theorem is not applicable in the case of intermediate compactification of type $M^{(5)} \times S^{1} \times K^{(4)} \to M^{(5)}$ when non-zero eigenvalues of Dirac operator on $K^{(4)}$ play the role similar to the bulk fermion masses in lower dimensions, that is they, as it was noted above, determine the bulk profiles of Fermi fields but do not prevent the appearance of chiral fermions in 4 dimensions. We’ll demonstrate how it works on the simplest example $K^{(4)} = S^{4}$.

The picture considered in this paper is a toy model that ignores the $SU(3) \times SU(2) \times U(1)$ group nature of the Standard Model. We also consider only the down-type quarks ($d, s, b$) that differ in $U(1)$ charge associated with the rotational symmetry around the brane. Thus the results of this paper are in a sense complementary to approach of \cite{Dolan1} where only single generation of the Standard Model spectrum is considered in a high-dimensional model incorporating the full group nature of Standard Model and four types of the SM fermions.

\section{Ten-dimensional background}

\quad  Let us begin with the action in $D$ dimensions for gravity described by metric $g_{AB}$ ($A, B = 0,1,2...(D-1)$), for scalar field $\phi$, $n$-form $F_{(n)i_{1}i_{2}... i{n}}$ and with $\Lambda$-term, $F_{(n)}$ and $\Lambda$ interacting with $\phi$:

\begin{equation}
\label{1}
S^{(D)} = M_{(D)}^{D-2} \int \left\{ R^{(D)} - \frac{1}{2} (\nabla\phi)^{2} - \frac{e^{\alpha\phi}}{2 \, n!} F_{(n)}^{2} - 2\Lambda e^{-\frac{\alpha}{n-1}\phi}\right\}\,\sqrt{-g^{(D)}} d^{D}x,
\end{equation}
where $M_{(D)}$ and $R^{(D)}$ are Planck mass and scalar curvature in $D$ dimensions. Following \cite{Duff1} we postulate the invariance of the action (\ref{1}) under the simultaneous scale transformation $g_{AB} \to e^{2\lambda} g_{AB}$, $\phi \to \phi + (2(n-1)/\alpha)\lambda$, $M_{(D)} \to e^{-\lambda} M_{(D)}$ ($\lambda = const$); this gives the coupling constant of $\phi$ in $\Lambda$-term in (\ref{1}). This Action is a modified and reduced version of the written in Einstein frame bosonic part of the Action of IIA supergravity (where $D = 10$, $n = 4$, $\alpha = 1/2$, $\Lambda = 0$, but plus to 4-form there are non-zero 2- and 3-forms \cite{IIA}).

The long time known \cite{Duff1}, \cite{Duff2} - \cite{Alt2} $p$-brane (fluxbrane) throat-like solution of the dynamical equations following from action (\ref{1}) looks in the deep of the throat as:

\begin{equation}
\label{2}
\begin{array}{c} 
ds_{(D)}^{2} = \left(\frac{r}{L}\right)^{2\beta (n-1)} \, \left[\eta_{\mu\nu}dx^{\mu}dx^{\nu} + U(r)\,\left(\frac{T_{\theta}}{2\pi}\right)^{2} d\theta^{2}\right] + \\
\\
+ \left(\frac{L}{r}\right)^{2\xi (n-1)} \, \left[\frac{dr^{2}}{U(r)} + \delta^{2} r^{2} d\Omega_{(n)}^{2}\right],
\end{array} 
\end{equation}

\begin{equation}
\label{3}
 e^{\phi} = e^{\phi_{0}} \left(\frac{r}{L}\right)^{\frac{2\alpha (n-1)}{\Delta}}, \,  F_{(n)} = Q_{(n)} dy_{1}\wedge... \wedge dy_{n}, \,  U(r) = 1 - \left(\frac{r_{Sch}}{r}\right)^{n-1},
\end{equation}
 where $\eta_{\mu\nu}$ is the most plus metric of the $p$-dimensional Minkowski space-time $M_{(p)}$; $\mu, \nu = 0, 1,... (p-1)$; $T_{\theta}$ is period of compact coordinate $S^{1}$ ($0 < \theta < 2\pi$); $r$ is isotropic coordinate along the throat; $d\Omega_{(n)}^{2}$ is volume element on sphere $S^{n}$ of unit radius, $y_{i}$ are angles of this sphere ($i = 1, 2,... n$); length $L$ and dimensionless constant $\delta$ are expressed through $Q_{(n)}$, $e^{\phi_{0}}$, $\Lambda$; $D = p + n +2$,

\begin{equation}
\label{4}
\begin{array}{c} 
\beta = \frac{2(n-1)}{(p+n)\Delta}, \, \,  \, \,  \,  \, \xi = \frac{2(p+1)}{(p+n)\Delta}, \, \,  \, \,  \,  \,  \Delta = \alpha^{2} + \frac{2(p+1)(n - 1)}{p+n}. 
\end{array} 
\end{equation}

In metric (\ref{2}) $\delta = 1$ if $\Lambda = 0$ in action (\ref{1}), non-zero $\Lambda$-term in (\ref{1}) results in $\delta \ne 1$, thus $\delta$  is a free parameter of the considered model, its physical meaning is clarified below.

The Schwarzschild type modification of solution when $U \ne 1$ in (\ref{2}) was considered in \cite{Duff2} - \cite{Grojean} in its Lorentzian version; here we use the Euclidean ”time” version of this deformation.

In order to avoid unnecessary cluttering of formulas, we will consider in the future only physically interesting values of dimensions $D = 10$, $p=4$, $n=4$ when, according to (\ref{3}), (\ref{4}):
 
\begin{equation}
\label{5}
\beta = \frac{3}{4\alpha^{2} + 15}, \qquad \xi = \frac{5}{4\alpha^{2} + 15}, \qquad U = 1 - \left(\frac{r_{Sch}}{r}\right)^{3}.
\end{equation}

Let us now perform the coordinate transformation 

$$
r \sim z^{- (15 + 4\alpha^{2})/(9 - 4\alpha^{2})}
$$
which takes metric (\ref{2}) (with parameters (\ref{5})) to the more familiar Poincare-like form, convenient for writing down the Dirac equation on this background:

\begin{equation}
\label{6}
ds_{(10)}^{2} = \frac{1}{(kz)^{2s}}\left[\eta_{\mu\nu}dx^{\mu}dx^{\nu} + U(z)\,\left(\frac{T_{\theta}}{2\pi}\right)^{2} d\theta^{2} + \frac{dz^{2}}{U(z)} + \kappa^{2} z^{2} d\Omega_{(4)}^{2}\right],
\end{equation}
where

\begin{equation}
\label{7}
U(z) = 1 - \left(\frac{z}{z_{IR}}\right)^{\gamma}, \quad  \gamma = \frac{3(15 + 4\alpha^{2})}{9 - 4\alpha^{2}}, \quad s = \frac{\gamma + 3}{8}, \quad \kappa = \delta \cdot \frac{3}{\gamma}.
\end{equation}
$k$ and $z_{IR}$ in (\ref{6}), (\ref{7}) are simply expressed through $L$ and $r_{Sch}$ in (\ref{2}), (\ref{3}), there is no need to put down these expressions since in what follows only background (\ref{6}), (\ref{7}) with its free parameters: period $T_{\theta}$ and dimensionless $\gamma$ and $\kappa$ will be used. Also, for completeness, we present the expression of the scalar field in terms of $z$ and $\gamma$:  $e^{\alpha\phi} \sim z^{3(5 - \gamma)/4}$, although this will not be needed in what follows.

We note that for $\alpha = 0$  in action (\ref{1}) metric (\ref{6}) describes the Schwarzschild deformed $AdS_{6} \times S^{4}$ space-time where $\kappa$ is the ratio of radius of $S^{4}$ to radius $k^{-1}$ of $AdS_{6}$. For $\alpha = 1/2$ in (\ref{1}) background (\ref{6}) is the Schwarzschild deformed IIA supergravity metric.

To determine finally the background it is necessary to match $z_{IR}$ and $T_{\theta}$. Following \cite{Duff2} - \cite{Alt2} we change coordinate $z$ in (\ref{6}), (\ref{7}) to $\tau$:

\begin{equation}
\label{8}
 z = z_{IR}\left[1 - \frac{\gamma}{4}\, \left(\frac{\tau}{z_{IR}}\right)^{2}\right], 
\end{equation}
which gives in vicinity of $z_{IR}$, $(z_{IR} - z) \ll z_{IR}$, for metric (\ref{6}) and for $U$ (\ref{7}): 

\begin{equation}
\label{9}
\begin{array}{c}
ds_{(10)}^{2} = \frac{1}{(kz_{IR})^{2s}}\left[\eta_{\mu\nu}dx^{\mu}dx^{\nu} + \eta^{2}\tau^{2}\,d\theta^{2} + d\tau^{2} + \kappa^{2} z_{IR}^{2} d\Omega_{(4)}^{2} \right], \\
\\
\eta = \gamma \cdot \frac{T_{\theta}}{4\pi z_{IR}}, \qquad U = \frac{\gamma^{2}}{4} \, \left(\frac{\tau}{z_{IR}}\right)^{2}.
\end{array}
\end{equation}
For $\eta = 1$ we have the smooth IR-end of the throat, for $\eta \ne 1$ there is conical singularity with the codimension two IR-brane at this point. As it was shown in \cite{Apple} in $6D$ model, in case $2 < \eta < 4$  there are three fermion generations corresponding to the quantum number along angular $\theta$ coordinate. Thus $\eta$ (\ref{9}) is one more, plus to $\gamma$ and $\kappa$ in (\ref{7}), free dimensionless parameter of the proposed model.

Throat (\ref{6}) is terminated at its IR end at $z = z_{IR}$. According to the conventional Randall-Sundrum approach we limit the throat (\ref{6}) also from below with the codimension one Planck UV-brane located at $z = z_{UV} = 1/k$, so that: 

\begin{equation}
\label{10}
z_{UV} = \frac{1}{k} \le z \le z_{IR}.
\end{equation}

Again, according to the familiar Randall-Sundrum approach, the Planck-TeV hierarchy ($\epsilon = 10^{-16}$) is equal to the ratio of warp factors at the UV and IR ends of the throat; for metric (\ref{6}) this means:

\begin{equation}
\label{11}
\epsilon = 10^{-16} = \left(\frac{z_{UV}}{z_{IR}}\right)^{s} = (k z_{IR})^{-s}.
\end{equation}

To conclude this section, a few remarks about UV and IR branes named above. At the UV end the extrinsic curvature, that is logarithmic derivative over $z$ of scale factors of subspaces $M_{(4)}$, $S^{1}$ and $S^{4}$ of metric (\ref{6}), and also derivative of scalar field $\phi(z)$ are non-zero. Hence $Z_{2}$ symmetry and corresponding Israel junction conditions on the codimension one Planck brane $M_{(4)} \otimes S^{1} \otimes S^{4}$ demand anisotropy of the UV-brane energy-momentum tensor. It is always possible to introduce the brane action which dependence on space-time metric and on scalar field ensures the fulfillment of the UV jump conditions. A more detailed discussion of this topic is beyond the scope of the present paper.

The case of the codimension two brane $M_{(4)} \otimes S^{4}$ at the IR end of the throat (\ref{6}) is theoretically less trivial \cite{Navarro1} - \cite{cod2}, while in our case it is technically more simple since there are no discontinuities on the brane in the extrinsic curvatures of brane's subspaces, and in the derivative of scalar field, which is evident if metric (\ref{9}) is considered near $z = z_{IR}$. This in turn means that action of the codimension two IR-brane cannot depend on the field $\phi$ and is of the isotropic Nambu-Goto type $S^{br}_{IR} = - M^{8}_{(10)} \oint \sigma \sqrt{-h}d^{8}x$,  where $h$ is determinant of the induced metric, $\sigma$ is dimensionless brane's tension. Then non-zero components of the brane's energy-momentum tensor are equal to: $T^{br \nu}_{\mu} \sim \sigma\delta^{\nu}_{\mu}$, $T^{br i}_{k} \sim \sigma\delta^{i}_{k}$, and $T^{br} \sim 8 \sigma$ ($T^{br}$ is trace of the brane's energy momentum tensor).

Although components $T^{br}_{AB}$ are proportional to delta function fixing the position of IR-brane in 10 dimensions ($\delta (\tau)/[2\pi \sqrt{g_{\tau\tau} g_{\theta\theta}}]$ for metric (\ref{9})), the components $(\mu\nu)$ and $(ik)$ of the RHS of the Einstein equations written in the form $M^{8}_{(10)}R^{(10)}_{AB} = T^{br}_{AB} - \frac{1}{8} g_{AB} T^{br}$  identically vanish. Hence there is no delta function in the components $R^{\mu}_{\nu}$, $R^{i}_{k}$ of the Ricci tensor and, as expected, no discontinuity in corresponding extrinsic curvatures. 

On the other hand the scaling factor before the $S^{1}$ subspace of metric (\ref{9}) is singular at $\tau = 0$, its logarithmic derivative is equal to $1/ \tau$. The terms of the $\theta\theta$ component of the Einstein equations that include the delta function $\delta(\tau)$ give the known relation between brane tension and angle deficit factor: $\sigma = 2\pi (1 - \eta)$ \cite{Navarro1} - \cite{cod2}. Hence the choice of arbitrary parameter $\eta$ (\ref{9}) of the model  is equivalent to the choice of the IR-brane dimensionless tension $\sigma$.

In summarizing this section, it is worth to emphasize that ten-dimensional background of the model under consideration depends on three dimensionless parameters: $\kappa$ in (\ref{6}), $\gamma$ in potential $U(z)$ (\ref{7}), and angle deficit factor $(1 - \eta)$ of the conical sungularity at the Schwarzschild Euclidean "horizon" $U(z_{IR}) = 0$ where the value of $\eta$ is given by ratio $T_{\theta} / z_{IR}$ (\ref{9}). The values of the calculated fermion masses depend also on the choice of $z_{IR}^{-1}$ (conventionally taken above several TeV), on the ratio $z_{UV} / z_{IR}$ (\ref{11}), and on some other constants.

But we will be interested here in the ratios of fermion masses which depend solely on three dimensionless parameters named above. Physically interesting results of Sec. 6 are obtained at the following integer values of these parameters: $\kappa = 3$, $\gamma = 9$ (that is $\alpha = \sqrt{3}/2$ in primary action (\ref{1}), see (\ref{7})), and $\eta = 3$ (see (\ref{24}) below).

\section{Dirac equation}

\quad Dirac equation with zero bulk mass in ten dimensions follows from the action

\begin{equation}
\label{12}
S_{\Psi} = \int {\bar\Psi}_{(32)} {\tilde\Gamma}^{A}D_{A}\Psi_{(32)} \sqrt{-g_{(10)}}d^{10}x,
\end{equation}
where ${\tilde\Gamma}^{A}$ are $32 \times 32$ gamma matrices in curved 10-dimensional space-time, $D_{A}$ are covariant derivatives with account of spin connection. We first write down Dirac equation for the general metric of type (\ref{6})

\begin{equation}
\label{13}
ds_{(10)}^{2} = b^{2}(z)\eta_{\mu\nu}dx^{\mu}dx^{\nu} + P^{2}(z) d\theta^{2} + N^{2}(z)dz^{2} + a^{2}(z) d\Omega_{(4)}^{2},
\end{equation}
making the following choice of flat anti-commuting $\Gamma^{A}$ (cf. \cite{Arutyunov}, \cite{Alt3}):

\begin{equation}
\label{14}
\begin{array}{c}
\Gamma^{\mu} = \gamma^{\mu} \otimes \sigma^{1} \otimes I_{4}; \qquad  \Gamma^{z} = \gamma_{5} \otimes \sigma^{1} \otimes I_{4}; \\
\Gamma^{\theta} = I_{4} \otimes \sigma^{2} \otimes \tau_{5}; \qquad \Gamma^{i} = I_{4} \otimes \sigma^{2} \otimes \tau^{i},
\end{array}
\end{equation}
where $\gamma^{\mu}$ are ordinary gamma matrices in Minkowski space-time, $\tau^{i}$ are the same on $S^{4}$, $\gamma_{5} = i\gamma^{0}\gamma^{1}\gamma^{2}\gamma^{3}$, $\tau_{5} = \tau^{1}\tau^{2}\tau^{3}\tau^{4}$, $\sigma^{1,2}$ are Pauli matrices, $I_{4}$ is $4 \times 4$ unit matrix. Corresponding Dirac equation on space-time (\ref{13}) (prime means derivative over $z$):

\begin{equation}
\label{15}
\begin{array}{c}
\left[\frac{1}{b}\Gamma^{\mu}\frac{\partial}{\partial x^{\mu}} + \frac{1}{N}\Gamma^{z}\left(\frac{\partial}{\partial z} + 2\frac{b^{\prime}}{b} + \frac{1}{2}\frac{P^{\prime}}{P} + 2\frac{a^{\prime}}{a}\right) + \frac{1}{P}\Gamma^{\theta} \frac{\partial}{\partial\theta} + \frac{1}{a}\Gamma^{i}\nabla_{i}\right]\Psi_{(32)} = 0.
\end{array}
\end{equation}

32-component spinor may be presented as a product of 8-component spinor $\Psi_{(8)}$ in 6-dimensional space-time $\{x^{\mu}, \theta, z\}$ and some eigenfunction $\chi_{l}(y^{i})$ of Dirac operator $\Gamma^{i}\nabla_{i}$ on sphere $S^{4}$ of unit radius: 

\begin{equation}
\label{16}
\Psi^{(l)}_{(32)} = \Psi^{(l)}_{(8)} (x^{\mu}, \theta, z) \cdot \chi_{l}(y^{i}); \quad \Gamma^{i}\nabla_{i} \chi_{l} = i K^{(4)}_{l}\chi_{l}; \quad K^{(4)}_{l} = \pm (l + 2),
\end{equation}
$l = 0, 1, 2...$ (eigenvalues of Dirac operator on sphere $S^{n}$ are: $K^{(n)}_{l} = \pm(l + n/2)$ \cite{Camporesi}).

8-component spinor is in turn a couple of 4-component spinors $(\pm)$, each of them consisting of right and left Weyl 2-spinors. Thus for definite eigenvalue of Dirac operator on $S^{4}$ and definite angular mode $q$ ($q = 0, 1, 2...$):

\begin{equation}
\label{17}
\Psi^{(l,q)}_{(8)} = \left(\begin{array}{ccc}
\psi_{R}^{+}(x)F_{R}^{+}(z) \\
\psi_{L}^{+}(x)F_{L}^{+}(z) \\
\psi_{R}^{-}(x)F_{R}^{-}(z) \\
\psi_{L}^{-}(x)F_{L}^{-}(z)
\end{array}\right) \cdot \frac{e^{i q \theta}}{b^{2}P^{\frac{1}{2}}a^{2}},
\end{equation}
here and below we omit the indexes $(l,q)$ of $\psi^{\pm}_{R,L}$ and profiles $F^{\pm}_{R,L}$.

From the Dirac equations $(\gamma^{\mu}\partial_{\mu} - m^{\pm})\psi^{\pm}(x) = 0$ for 4-spinors $\psi^{\pm} = \psi^{\pm}_{R} + \psi^{\pm}_{L}$ ($m^{\pm}$ are the fermion masses in four dimensions), and with account of (\ref{16}), (\ref{17}) Dirac equation (\ref{15}) comes to four equations for profiles $F^{\pm}_{R,L}$:

\begin{equation}
\label{18}
\left\{\begin{array}{c}
\left(\frac{1}{N}\frac{d}{dz} + \frac{q}{P} + \frac{K^{(4)}_{l}}{a}\right) F^{-}_{R} - \frac{m^{-}}{b} F^{-}_{L} = 0, \\
\\
\left(\frac{1}{N}\frac{d}{dz} - \frac{q}{P} - \frac{K^{(4)}_{l}}{a}\right) F^{-}_{L} + \frac{m^{-}}{b} F^{-}_{R} = 0,
\end{array}\right.
\end{equation}

\begin{equation}
\label{19}
\left\{\begin{array}{c}
\left(\frac{1}{N}\frac{d}{dz} - \frac{q}{P} - \frac{K^{(4)}_{l}}{a}\right) F^{+}_{R} - \frac{m^{+}}{b} F^{+}_{L} = 0, \\
\\
\left(\frac{1}{N}\frac{d}{dz} + \frac{q}{P} + \frac{K^{(4)}_{l}}{a}\right) F^{+}_{L} + \frac{m^{+}}{b} F^{+}_{R} = 0.
\end{array}\right.
\end{equation}
We note that equations for ($+$) and ($-$) components of $6D$ 8-spinor separate thanks to the choice (\ref{14}) of higher-dimensional gamma-matrices.

Finally, taking the scale factors $b(z), P(z), N(z), a(z)$ from comparison of metrics (\ref{13}) and (\ref{6}), multiplying equations (\ref{18}), (\ref{19}) by $N$, defining constants $c_{l}$ which are analogous to $c_{f} = M_{f}^{bulk} / k$ in $5D$ models of \cite{5D1} - \cite{5D3}, and passing to dimensionless quantities $t$, $\mu^{\pm}$:

\begin{equation}
\label{20}
t = \frac{2\pi z}{T_{\theta}}; \quad \frac{\gamma}{2\eta}\, \epsilon^{\frac{1}{s}} = t_{UV} < t < t_{IR} = \frac{\gamma}{2\eta};  \quad \mu^{\pm} = \frac{T_{\theta}}{2 \pi} \cdot m^{\pm}; \quad c_{l} = \frac{K^{(4)}_{l}}{\kappa},
\end{equation}
(the lower and upper limits for the variable $t$ are determined from (\ref{9}) - (\ref{11}), $\epsilon = 10^{-16}$) we obtain from (\ref{18}) the following system of equations for the profiles $F^{-}_{R,L}(t)$:

\begin{equation}
\label{21}
\left\{\begin{array}{c}
\left(\frac{d}{dt} + \frac{q}{U} + \frac{c_{l}}{t \, \sqrt{U}}\right) F^{-}_{R} - \frac{\mu^{-}}{\sqrt{U}}F^{-}_{L} = 0,\\
\\
\left(\frac{d}{dt} - \frac{q}{U} - \frac{c_{l}}{t \, \sqrt{U}}\right) F^{-}_{L} + \frac{\mu^{-}}{\sqrt{U}}F^{-}_{R} = 0.
\end{array}\right. \quad U = 1 - \left(\frac{2\eta\, t}{\gamma}\right)^{\gamma}
\end{equation}
Equations (\ref{19}) for $F^{+}_{R,L}$ take a similar form with corresponding changes in the signs of various terms. It is important to note that profiles of zero modes of $F^{-}_{R}$, $F^{+}_{L}$ and of $F^{-}_{L}$, $F^{+}_{R}$ coincide which is easily seen from (\ref{18}), (\ref{19}) in case $m^{\pm} = 0$; this will be used in Sec. 6.

\section{Three generations from one Fermi field}

\quad The coefficient in action (\ref{12}) at the kinetic term in 4 dimensions ${\bar\psi}\gamma^{\mu}\partial_{\mu}\psi$ must be equal to one for each mode $q$, $l$, $\pm$ and separately for $R$ and $L$ Weyl components. Thus, with account of $\sqrt{g_{(10)}} = b^{4} P N a^{4}$ (we again use generic metric (\ref{13})), $\Psi^{2}_{(8)} \sim (b^{4} P a^{4})^{-1}$ (see (\ref{17})), $\int {\bar\chi}_{l}\chi_{l^{\prime}} d\Omega_{(4)} = \delta_{l\,l^{\prime}}$, $\int e^{i(q - q^{\prime})\theta} d\theta = 2\pi \delta_{q\,q^{\prime}}$ the following normalization condition of functions $F^{\pm}_{R,L}(z)$ must be satisfied:

\begin{equation}
\label{22}
2\pi \int_{z_{UV}}^{z_{IR}} \frac{N}{b} (F^{\pm}_{R,L})^{2} dz = 2\pi \int_{z_{UV}}^{z_{IR}} \frac{1}{\sqrt{U}} (F^{\pm}_{R,L})^{2} dz = 1.
\end{equation}

Transforming in this integral and in equations (\ref{18}), (\ref{19}) coordinate $z \to \tau$ like in (\ref{8}), and taking into account that near the $IR$ end of the throat $F_{R,L} \sim \tau^{\pm q/\eta}$ (see (\ref{18}), (\ref{19})) and $dz / \sqrt{U} \sim d\tau$ (see expression for $U$ in (\ref{9})) we see that finitness of integral (\ref{22}) demands:

\begin{equation}
\label{23}
\int_{0} \tau^{\pm \frac{2q}{\eta}} d\tau  < \infty.
\end{equation}
For smooth $IR$ end of metric (\ref{6}) or (\ref{9}) ($\eta = 1$) this integral is non-divergent only for one mode $q = 0$. In case

\begin{equation}
\label{24}
2 < \eta < 4
\end{equation}
integral (\ref{23}) is finite for three modes (three fermion generations) $q = 0, \pm 1$. Thus, we reproduce here result of \cite{Apple} but not for the extra 2-sphere as a background but for the Schwarzschild deformed supergravity background (\ref{6}), (\ref{7}). Below it is shown that physically interesting results are obtained for $\eta = 3$ in (\ref{9}).

\section{Twisted BC: 3 small masses}

\quad In what follows we set $U = 1$ in system (\ref{21}) for $F^{-}_{R,L}$ and in similar equations for $F^{+}_{L,R}$. As it is discussed in the end of Sec. 6, taking into account $U \ne 1$, as in (\ref{7}), (\ref{21}), only slightly changes the calculated mass ratios of quarks of different generations.

For $U = 1$ and $q = 0$ system (\ref{21}) coincides with equations in $5D$ models \cite{5D1} - \cite{5D3} for one Fermi field on a slice (\ref{10}) of $AdS_{5}$ with an essential difference, already noted in the Introduction, that in $5D$ models the values of parameters $c_{f} = M^{bulk}_{f} / k$ are arbitrary for every fermion whereas in (\ref{21}) $c_{l}$ are geometrically motivated (see (\ref{20})). For $q \ne 0$ the solutions of equations (\ref{21}) differ in an interesting way from those in the models \cite{5D1} - \cite{5D3}, this is what we will discuss now.

For $q \ne 0$, $m^{-} \ne 0$ in (\ref{18}) (or equivalently $\mu^{-} \ne 0$ in (\ref{21})), the following second-order equations for $F^{-}_{R,L}$ are obtained from (21):

\begin{equation}
\label{25}
\begin{array}{c}
\frac{d^{2}F^{-}_{R}}{d t^{2}} - \left[\omega^{2} + \frac{2qc_{l}}{t} + \frac{c_{l}(c_{l} + 1)}{t^{2}}\right] F^{-}_{R} = 0, \\
\frac{d^{2}F^{-}_{L}}{d t^{2}} - \left[\omega^{2} + \frac{2qc_{l}}{t} + \frac{c_{l}(c_{l} - 1)}{t^{2}}\right] F^{-}_{L} = 0,
\end{array} \qquad \omega = \sqrt{q^{2} - (\mu^{-})^{2}}.
\end{equation}

It is curious to note that these equations coincide with the non-relativistic Schrödinger equation for an electron moving in a Coulomb field, where $c_{l}$ in (\ref{25}) play the role of the electron's orbital momentum. The solutions of these equations, as it is well known \cite{Landau}, are expressed through the degenerate hypergeometric functions $\Phi(\alpha, \gamma; \rho)$. Correspondingly for solutions of Eq-s (\ref{25}), with account of the first order connections (\ref{21}) of $F^{-}_{R}$ and $F^{-}_{L}$, it is obtained ($A, B$ are two arbitrary constants):

\begin{equation}
\label{26}
\begin{array}{c}
F^{-}_{R} = A \, \frac{\mu^{-}}{2\omega}\, \frac{1}{2 c_{l} + 1} \, \rho^{c_{l} + 1} \, e^{- \frac{\rho}{2}} \, \Phi\left(c_{l} + \frac{qc_{l}}{\omega} + 1, 2c_{l} + 2; \rho\right) + \\
\\
+ B\, \, \rho^{- c_{l}} \, e^{- \frac{\rho}{2}} \, \Phi\left(- c_{l} + \frac{qc_{l}}{\omega}, - 2c_{l}; \rho\right),
\end{array} \qquad \rho = 2 \, \omega\, t
\end{equation}

\begin{equation}
\label{27}
\begin{array}{c}
F^{-}_{L} = A \, \rho^{c_{l}} \, e^{- \frac{\rho}{2}} \, \Phi\left(c_{l} + \frac{qc_{l}}{\omega}, 2c_{l}; \rho\right) + \\
\\
+ B\, \frac{\mu^{-}}{2\omega}\, \frac{1}{2 c_{l} - 1} \, \rho^{- c_{l} + 1} \, e^{- \frac{\rho}{2}} \, \Phi\left(- c_{l} + \frac{qc_{l}}{\omega} + 1, - 2c_{l} + 2; \rho\right).
\end{array} \qquad \rho = 2 \, \omega\, t
\end{equation}
Solutions for $F^{+}_{R,L}$ differ from the presented solutions for $F^{-}_{R,L}$ by the changing of signs of $q$ and $c_{l}$ in (\ref{26}), (\ref{27}). 

For $q = 0$ ($\omega = i\mu^{-}$ in (\ref{25}) - (\ref{27})) solutions of Eq-s (\ref{21}) are the familiar expressions in Bessel functions \cite{5D1} - \cite{5D3} ($A$, $B$ are another arbitrary constants):

\begin{equation}
\label{28}
\begin{array}{c}
F^{-}_{R} = A \, \sqrt{t} \, J_{c_{l} + \frac{1}{2}} (\mu^{-}t) + B\, \sqrt{t} \, J_{-(c_{l} + \frac{1}{2})} (\mu^{-} t), \\
\\
F^{-}_{L} = A \, \sqrt{t} \, J_{c_{l} - \frac{1}{2}} (\mu^{-}t) - B\, \sqrt{t} \, J_{-(c_{l} - \frac{1}{2})} (\mu^{-} t).
\end{array}
\end{equation}

It is known that Hermiticity of Dirac operator in (\ref{18}) (or (\ref{21})) requires the fulfillment of the boundary conditions (BC):

\begin{equation}
\label{29}
F^{-}_{L}(z_{UV}) \, F^{-}_{R}(z_{UV}) = F^{-}_{L}(z_{IR}) \, F^{-}_{R}(z_{IR}) = 0.
\end{equation}
For zero modes, $m^{-} = 0$ in (\ref{18}), this means: or $F^{-}_{L} \equiv 0$, or $F^{-}_{R} \equiv 0$; we'll discuss the zero modes option in the next Section. In turn, BC $F^{-}_{L} (z_{UV}) = F^{-}_{L}(z_{IR}) = 0$ (or $F^{-}_{R} (z_{UV}) = F^{-}_{R}(z_{IR}) = 0$) give a tower of Kaluza-Klein (KK) heavy modes with masses $m^{-}_{n}$, their lower value is of order $z^{-1}_{IR}$. To escape the observationally forbidden phenomena, like flavor changing neutral currents generated by the exchange of these KK heavy modes, $z_{IR}^{-1}$ must be above several TeV. Thus heavy modes can not be the low mass quarks or leptons.

In the next section, the commonly used and corresponding to the Standard Model ideology  method of obtaining small quarks' current masses in higher dimensional theories by introducing a Higgs field located on or near the IR brane will be applied. But now we'll consider the appearance of small fermion masses by imposing on solutions (\ref{26}) - (\ref{28}) the so called "twisted" BC, asymmetric on the two end points, first proposed in \cite{twisted1} (see also \cite{Alt3}, \cite{twisted2}):

\begin{equation}
\label{30}
F^{-}_{L}(z_{UV}) = F^{-}_{R}(z_{IR}) = 0.
\end{equation}
This BC excludes zero mode $m = 0$, and it gives of course its tower of heavy modes $m_{n} > z^{-1}_{IR}$. But twisted BC (\ref{29}) also gives the single (for every $q$, $l$ in (\ref{18}), (\ref{19}) or (\ref{21})) mode with small mass $m_{q,l}^{tw}$ for which $m_{q,l}^{tw} z \ll 1$ everywhere on the interval (\ref{10}).

For $q = 0$ twisted BC gives (received from (\ref{28}) with account that $\mu^{tw} t \ll 1$):

\begin{equation}
\label{31}
m^{tw}_{q=0,l} = \frac{1}{z_{IR}}\, \sqrt{4c^{2}_{l} - 1} \, \left(\frac{z_{UV}}{z_{IR}}\right)^{c_{l} - \frac{1}{2}}, \qquad c_{l} > \frac{1}{2},
\end{equation}
here $z_{UV} / z_{IR} \ll 1$ (see (\ref{11})), from now on we'll consider physically interesting positive values of $c_{l}$ (\ref{20}) with values above $1/2$, like $c_{f}$ in \cite{5D1} - \cite{5D3}. Thus fermion mass (\ref{30}) is really small, $m^{tw} \ll z^{-1}_{IR}$.

For $q = \pm 1$ expanding $\omega$ (\ref{25}) over $\mu^{-}$ in case $q \ne 0$: $\omega = |q| [1 - (\mu^{-})^{2}/ 2q^{2}]$, and applying it in solutions (\ref{26}), (\ref{27}) in the lowest order in $\mu^{-}$ it is obtained from BC (\ref{30}):
 
\begin{equation}
\label{32}
m^{tw}_{q = 1,l} = m^{tw}_{q=0,l} \cdot \frac{1}{\sqrt{\Phi\left(2c_{l} + 1, 2c_{l} + 2; \frac{\gamma}{\eta}\right)}},
\end{equation}

\begin{equation}
\label{33}
m^{tw}_{q = -1,l} = m^{tw}_{q=0,l} \cdot \frac{1}{\sqrt{\Phi\left(2c_{l} + 1, 2c_{l} + 2; - \frac{\gamma}{\eta}\right)}},
\end{equation}
$m^{tw}_{q=0,l}$ is given in (\ref{31}) and the upper value of variable $t = t_{IR} = \gamma/2\eta$, see (\ref{20}), was used here with account that argument of the degenerate hypergeometric functions in (\ref{26}), (\ref{27}) $\rho_{max} = 2\omega t_{IR} = 2 |q| t_{IR} = \gamma/\eta$ (for $|q| = 1$).

It may be shown that for $c_{l} > 1/2$, $\gamma / \eta > 1$ masses (\ref{31}) - (\ref{33}) obey: $m^{tw}_{q = 1,l} < m^{tw}_{q=0,l} < m^{tw}_{q = -1,l}$, thus, for example in case $l = 0$, it would be natural to try to identify these masses, say, with masses of $(d, s, b)$ quarks which experimentally observed ratios \cite{masses}

\begin{equation}
\label{34}
\frac{m_{d}}{m_{s}} = 5.0(7) \cdot 10^{-2}; \qquad  \qquad \frac{m_{s}}{m_{b}} = 2.22(25) \cdot 10^{-2}.
\end{equation}
are the main focus of this paper.

Unfortunately, the mass ratios given by the expressions (\ref{32}), (\ref{33}) do not agree with the experimental data (\ref{34}) for any values of the arguments $c_{l} > 1/2$, $\gamma / \eta > 1$ of the degenerate hypergeometric functions in the RHS of (\ref{32}), (\ref{33}). The conventional Higgs mechanism for obtaining small fermionic masses discussed in the next Section proves to be more successful, although not perfect as well, in frames of the considered toy model.

\section{Higgs mechanism and zero modes profiles}

\quad For $m^{\pm} = 0$ in Eq-s (\ref{18}), (\ref{19}), or equivalently $\mu^{-} = 0$ in Eq-s (\ref{21}) for $F^{-}_{R.L}$ and $\mu^{+} = 0$ in similar equations for $F^{+}_{R,L}$, and for $U = 1$ in these equations we have four zero modes profiles 

\begin{equation}
\label{35}
\begin{array}{c}
F^{-}_{R} = C^{-}_{R} \, e^{-qt} \, t^{-c_{l}}, \qquad F^{-}_{L} = C^{-}_{L} \, e^{qt} \, t^{c_{l}},\\
\\
\qquad F^{+}_{R} = C^{+}_{R} \, e^{qt} \, t^{c_{l}}, \quad \qquad F^{+}_{L} = C^{+}_{L} \, e^{-qt} \, t^{-c_{l}}.
\end{array}
\end{equation}

Since Higgs field is located on or near the IR brane small fermion masses are acquired by the fields which profiles have maximum at the other - $UV$ end of the throat where $t \ll 1$ (see definition of $t$ in (\ref{20})). Thus, out of four solutions (\ref{35}) two proportional to $t^{-c_{l}}$ must be left (we remind that $c_{l} > 1/2$), while $C^{-}_{L} $ and $C^{+}_{R}$ are set equal to zero. Which in turn ensures the fulfillment of the boundary conditions (\ref{29}).

Nonzero constants $C^{-}_{R}$ and $C^{+}_{L}$ are equal to each other and are determined from the normalization integral (\ref{22}) which being rewritten through variable $t$ (\ref{20}), for $U = 1$, and after substitution non-zero profiles (\ref{35}) takes a form ($C$ is any of two named constants): $C^{2} T_{\theta} \int_{t_{UV}}^{t_{IR}}\, e^{-2qt}t^{-2c_{l}}dt = 1$. Since the main contribution to this integral comes from small $t$ where $qt \ll 1$ exponent in the integrand can be ignored and neglecting the small contribution from the upper limit of integration we come to the  normalization condition

\begin{equation}
\label{36}
(C^{-}_{R})^{2} = (C^{+}_{L})^{2} = \frac{2c_{l} - 1}{T_{\theta}} \, t^{2c_{l} - 1}_{UV},
\end{equation}
which dependence on $c_{l}$ is familiar from the similar conditions in $5D$ models \cite{5D1} - \cite{5D3}.

The conventional way to introduce the SM Yukawa interactions is to add to spinor action (\ref{12}) the Higgs field mass term:

\begin{equation}
\label{37}
S_{H} = \int {\bar\Psi}_{(32)} H \Psi_{(32)} \sqrt{-g_{(10)}}d^{10}x,
\end{equation}
where vacuum average of the Higgs field $\langle H \rangle$ "lives" in vicinity or at the IR end of the throat. After substitution here expressions for $\Psi_{(32)}$ from (\ref{16}), (\ref{17}) and integration over extra coordinates with assumption that Higgs field does not depend on $\theta$ snd $y^{i}$, the non-zero diagonal in $l, q$ mass terms of the fermion action in 4 dimensions $S_{m}^{(l,q)} = \int {\bar \psi}^{l,q}m_{l,q}\psi^{l,q} d^{4}x$ are obtained for every component $(l,q)$ of the high dimensional spinor $\Psi_{(32)}$, where

\begin{equation}
\label{38}
m_{l,q} = 2\pi \int F^{-(l,q)}_{R}\, F^{+(l,q)}_{L}\, \langle H \rangle N dz,
\end{equation}
$N$ is lapse function in metrics (\ref{13}) or (\ref{6}).

It makes sense to note that, in contrast to the Dirac equation (\ref{15}), where due to the choice of gamma matrices (\ref{14}), the equations for two 4-components of the 8-component spinor (\ref{17}) are separated (see (\ref{18}), (\ref{19})), the mass Higgs term of the action (\ref{37}) entangles these $\pm$ 4-spinors, providing the necessary product of right and left Weyl spinors - like in (\ref{38}). We believe that this is an important advantage of the model under consideration, since, for example, in the $5D$ models \cite{5D1} - \cite{5D3}, right and left spinors are introduced independently and are quite arbitrarily combined into the Higgs terms of the Lagrangian, giving Yukawa interactions.

Finally, under the assumption $\langle H \rangle = Y \, \delta(z -z_{IR}) / N$ ($Y$ is dimensionless Yukawa coupling constant), writing down from (\ref{35}), (\ref{36}) expressions for $F^{-}_{R} (t_{IR})$, $F^{+}_{L}(t_{IR})$, taking into account the relationships (\ref{9}), (\ref{20}) between $T_{\theta}$, $z_{IR}$, and $t_{IR} = \gamma / 2\eta$  we get from (\ref{38}):

\begin{equation}
\label{39}
m_{l,q} = Y \, \frac{2c_{l} - 1}{z_{IR}} \, \left(\frac{z_{UV}}{z_{IR}}\right)^{2c_{l} - 1} \, e^{-q \cdot \frac{\gamma}{\eta}}.
\end{equation}

In case the 4-dimensional compact subspace (the base) of 10-dimensional supergravity background is sphere $S^{4}$, like it is supposed in (6), we have $c_{l} = (l + 2) / \kappa$ (see (\ref{16}), (\ref{20})), where $\kappa$ in metric (\ref{6}) is a free parameter of the considered model. Assuming $\kappa = 3$ gives

\begin{equation}
\label{40}
c_{l=0} = \frac{2}{3}, \qquad  c_{l=1} = 1,
\end{equation}
and for $l = 0$ it is worth again to try to identify three masses (\ref{39}) obeying inequalities $m_{q = 1,l=0} < m_{q=0,l=0} < m_{q = -1,l=0}$ with masses of three down type quarks $(d, s, b)$.

For ratios of these masses (\ref{39}) gives:

\begin{equation}
\label{41}
\frac{m_{d}}{m_{s}} = \frac{m_{s}}{m_{b}} = e^{- \frac{\gamma}{\eta}}.
\end{equation}
Assuming, like it was pointed out in the end of Sec. 2, the following integer values for two other free parameters of the model: $\gamma = 9$ in potential $U(z)$ (\ref{7}) and $\eta = 3$ (introduced in (\ref{9}), its permitted values see in (\ref{24})), the observed value $5.0(7) \cdot 10^{-2}$ (\ref{34}) of ratio of masses of $d$ and $s$ quarks is obtained from (\ref{41}): $m_{d} / m_{s} = e^{-3} = 4.97 \cdot 10^{-2}$. Whereas the observed value (\ref{34}) of ratio $m_{s} / m_{b}$ is twice less than the one predicted in (\ref{41}). The equality of ratios of lighter and heavier quark masses in (\ref{41}) is a difficulty of the considered toy model. The possibility to overcome this difficulty is discussed in the Conclusion.

Let us look now at the the supposed mass of $b$ quark $m_{b} = m_{l=0, q= -1}$ (\ref{39}), when the assumed values of three parameters of the model ($\kappa = 3$, $\gamma = 9$, $\eta =3$) are used. Then $c_{0} = 2/3$ (see (\ref{40})) and from (\ref{11}) the ratio $z_{UV} / z_{IR} = 10^{-16/s} = 2.17 \, 10^{-11}$ ($s = 3/2$ for $\gamma = 9$, see (\ref{7})) is obtained. Finally from (\ref{39}) $m_{b} = Y\, z_{IR}^{-1} \cdot 2 \, 10^{-3}$. This corresponds to the experimentally observed value $m_(b) = 4.18(2)$ GeV \cite{masses} if dimensionless Yukawa coupling $Y$ is of the order of one and $z_{IR}^{-1}$ is equal to its conventionally accepted value about 1 TeV.

The repetition of similar calculations with the same input data for the case $l = 1$ in (\ref{39}), that is for $c_{l} = 1$ (see (\ref{40})), gives $m_{l=1, q= -1} \approx 10^{2}$ eV. Of course, it would be wonderful if the spectral number $l$ enumerated types of quarks, while $q$ corresponded to the generations of quarks of the same type. But this can hardly be expected in the framework of the considered extremely simplified model.

Now a few final remarks about the approximation $U = 1$ in Eq-s (\ref{21}) and, accordingly, in all further expressions.
It is interesting that in case $\mu^{-} \ne 0$ considered in Sec. 5 coordinate transformation (\ref{8}) in the exact (that is for $U \ne 1$) Eq-s (\ref{21}) gives in the vicinity of $z = z_{IR}$, when Dirac equation is written down on the rather simple background metric (\ref{9}), the approximate equations similar to (\ref{21}) with $U = 1$ where the roles of two parameters $q$, $c_{l}$ are reversed. Correspondingly, these parameters have to be swaped in Eq-s (\ref{25}) and in solutions (\ref{26}), (\ref{27}) of new equations. The conjugation of two approximate solutions, one being  valid near the IR end and another near the UV end of the throat, in some intermediate point may allow one to perform more accurate calculations of small masses (\ref{31}) - (\ref{33}). But this work hardly makes sense, given the unrealistic values of the small masses received from twisted BC.

As for the correction of zero-mode profiles (\ref{35}) of Sec. 6, then again it is not difficult to write down these profiles near $z = z_{IR}$, that is near $\tau = 0$, using the coordinate transformation (\ref{8}) and metric (\ref{9}).
 But now, in contrast to the situation considered above in Sec. 6, the zero-modes profiles are singular at $\tau = 0$ (although their norms are integrable for $q = 0, \pm 1$, see Sec. 4). Assumption $\langle H \rangle = Y \, \delta(z -z_{IR}) / N$ does not work now, to get the Higgs generated masses of type (\ref{38}), (\ref{39}) it is necessary to put $\langle H \rangle = const$ in some interval in vicinity of $z = z_{IR}$, like it was done, for example, in \cite{Neronov}. Estimates show that this way of taking into account $U \ne 1$ slightly changes the result (\ref{41}) for the quark mass ratios.

\section{Conclusion}

\quad Three main results of this paper may be outlined.

It is shown that using 10 dimensional supergravity backgrounds the usually arbitrarily selected in the models of 5D warped compactifications fermions' bulk masses may be identified with the eigenvalues of the Dirac operator on a compact 4D subspace $K^{(4)}$. This means in particular that Witten’s no-go theorem \cite{Witten} is not applicable in the case of intermediate compactification of type $M^{(5)} \times S^{1} \times K^{(4)} \to M^{(5)}$ and that in order to obtain chiral fermions in four dimensions it is not necessary to have chiral fermion modes on $K^{(4)}$, as it is done in 10D models of papers \cite{Dolan1} - \cite{Dvali} and in many other works. The most simple option $K^{4} = S^{4}$ considered in the paper is not a realistic one but perhaps the consideration of the non-chiral modes of more sophisticated compact subspaces $K^{(4)}$  with, generally speaking, topologically non-trivial background gauge fields will prove to be useful in the modeling rich SM structure with four types of fermions (up and down quarks, charged leptons and neutrinos) and the known SM group nature.

It is shown that Schwarzschild Euclidean deformation of the generalized IIA supergravity background with certain angle deficit factor of the conical singularity at the "horizon" leaves non-divergent three fermion angular modes interpreted as three generations of fermions of one and the same type. This result reproduces similar earlier results achieved perhaps in more artificial 6D models.

For $d$ and $s$ quarks the simple and experimentally exact expression for ratio of their masses $m_{d}/m_{s} = e^{-3}$ is obtained for the integer values of two free parameters of the model: exponent $\gamma = 9$ in Schwarzschild potential (\ref{7}) (this choice of $\gamma$ corresponds to the 4-form-dilaton coupling in primary Action (\ref{1}) $\alpha = \sqrt{3} / 2$), and angle deficit factor in (\ref{9}) $1 - \eta = - 2$. Although such an exact correspondence with the experiment for integer values of the free parameters of the model looks surprising, the problem of the geometric justification of the choice of precisely these values of parameters remains.

Experimentally ratios of masses of quarks of lighter and heavier generations differ essentially, whereas in the proposed approach they are equal to each other, like $m_{d} / m_{s} = m_{s} / m_{b}$ in (\ref{41}). The different dynamics for fermion modes with different angular momentum $q = 0, \pm 1$ corresponding to three generations is in demand as a remedy to this difficulty. This angular momentum $q$ is in fact a charge for the Kaluza-Klein gauge field associated with $U(1)$ symmetry of the $S^{1}$ subspace of the high dimensional space-time \cite{Neronov-2}. Introduction of such background Kaluza-Klein gauge field is perhaps the way to resolve the problem.

We did not discuss the possible origin of the Cabibbo-Kobayashi-Maskawa (CKM) matrix and the mixing of quark generations. This would make sense in the model incorporating up and down quarks which Yukawa matrices have non-diagonal elements. If Higgs field in (\ref{37}) depends on $\theta$ ($H \sim e^{ip\theta}$ \cite{Neronov} - \cite{Libanov}) then (\ref{37}) will generate not only diagonal elements (\ref{38}) of the Yukawa mass matrix but also its non-diagonal elements ($q \ne q^{\prime}$).  However, consideration of this possibility, including the calculation of the CKM matrix elements, is beyond the scope of the present paper. A surprising correlation (the so-called "flavor puzzle") between the experimentally observed values of quark masses and the CKM matrix mixing angles was considered in a recent author's paper \cite{Alt4}, which is not related to the high dimensional models.

We hope that results of the presented simplified model will contribute to the search of a more realistic high dimensional approach to the SM.

\section*{Acknowledgments} Author is grateful to Mikhail Vysotsky for constant friendly consultations and to Valery Rubakov, who tragically died in October 2022, for the inspiring discussions on the ABC of Standard Model and its problems.

\end{document}